\documentclass[twocolumn,showpacs,preprintnumbers,amsmath,amssymb]{revtex4}

\usepackage{graphicx}
\usepackage{dcolumn}
\usepackage{bm}

\begin{document}

\texttt{preprint}

\title{Stability Studies of Vortex State in Superconducting Disks}

\author{W.M. Wu},
\author{M.B. Sobnack},
\author{F.V. Kusmartsev}
\affiliation{
Department of Physics, Loughborough University, Loughborough LE11 3TU,
United Kingdom}
\date{\today}

\begin{abstract}
We are going to analyze the nucleation of vortices in a thin mesoscopic superconducting disk. The Gibbs free energy from London model is formulated. This energy function, with an arbitrary configuration of vortices, is assoicated with the disk's size, applied magnetic field and finite temperature. Then, the optimal solution is obtained by differentiating with respect position $r$, for fixed applied field and temperature. We also investigate the stability of the different vortex states inside the disk and compare our results with those of other theoretical studies and with available experimental observations. Our results agree with experiments. Besides, we formulate the Gibbs free energy by Ginzburg-Landau (GL) model. Theoretically, the free energy by GL model takes the superconducting density into account but not in London model. Our simulations from both theories show the same quantum states of vortex. We find that the Gibbs free energy by GL model is smaller than by London model.
\end{abstract}


\maketitle

\section{Introduction}
It is an interesting topic that how vortex nucleates inside mesocopic superfluids and superconductors for few decades. In our studies, we are going to study the vortices on the small disk. We formulate the Gibbs free energy in terms of the applied magnetic field $\bf {H}$, the size of sample and temperature. The London model and Ginzburg-Landau model are applied in our studies. The formulation of Gibbs free energies are combined with the kinetic energy, magnetic energy and superconducting density energy. We also modify the Gibbs free energies by adding the term of entropy. It means that the order-disorder of vortices inside the sample has been taken into account. We found that this Gibbs free energy by Ginzburg-Landau model can be aprroximated into the one by London model. Of course, the London theory only describes the macroscopic phenomena, while the GL considers the microscopic influences at the same time. After that, we will show that the GL model and London model agree with the experiments.

It is better to have an overview of some backgrounds of superconductors in mesoscopic scale. The nucleation of vortex in superfluid films and small superconductors as a function of flow velocity (superfluids) and applied magnetic field (superconductors) have been the subject in recent years. It is well known that the characteristics of small systems\cite{R1} are affected by the sizes and the geometries. As the size of the system becomes small, the boundary conditions become more important. In addition, the vortex interactions in mesoscopic scale samples are different from those in large samples. Geim and co-workers \cite{R2,R3,R4} found that there could be paramagnetic Meissner effect in small superconductors. Chibotaru {\it et al.}\cite{R5} and Mel'nikov \cite{R6} studied the possibility of anti-vortices and multi-quanta vortices penetrating thin mesoscopic square samples. Schweigert {\it et al.}\cite{R7,R8} concluded theoretically that the multi-vortex state would transform into a single giant vortex state as the magnetic field and the disk thickness increase. On the other hand, Okayasu and co-workers \cite{R9} reported that the giant vortex could not be explained according to their experiments. 

Most recent theoretical studies \cite{R10,R11,R12,R13,R14,R15} can explain the creation of different vortex states and their stability well. In our studies, we modify the current results of small thin disk and compare with experimental observations \cite{R16}. Experimentally, Grigorieva and co-workers \cite{R16} studied vortex configurations in mesoscopic superconducting disks, and found that the vortices form concentric rings or shells, rather like shell filling in atoms and nuclei. They found that the configrations of vortices (the so-called magic numbers), corresponding to the appearance of new shells, are reproducible in many experiments for different applied fields $H$ and diameter $D$ of their disks. There are some theories which explains the mechanism for vortex shell filling. However, we would like to modify the formulation of free energy by introducing the idea of entropy. Theories\cite{R15} also predict stable configurations well generally. Nonetheless, in some cases, they are different from those observed experiments \cite{R16}. It is appropriate to remark here that while experiments are performed at finite $T\neq 0$\,K temperature, most theoretical studies are only valid at $T=0$\,K.

Finally, we'll extend the preliminary work of Sobnack {\it et al.} \cite{R10} to include temperature by taking into account of the entropy energy associated with the vortex state. Using GL model and London model \cite {R17,R18,R19,R20,R21,R22,R23,R24}, we write down the free energies of the disk with an arbitrary configuration of vortices arranged in shells. The free energy is minimized and the optimal vortex configurations are then obtained. We also found that Gibbs free energy in GL model is less than from London model.

\section{Formulation and Methodology}
\subsection{Configuration of Mesoscopic Disk}
Assume that there are vortices inside a thin disk under the applied magnetic field $\bf H$ with radius $R$ and thickness $d$ \cite {R19,R20,R21,R22,R23,R24} (volume is $V=\pi R^2 d$). The small disk in an applied magnetic field is under ${\bf H}=H{\bf k}= \nabla \times {\bf A_{\rm app}}$, perpendicular to the plane of the disk. We restrict the scale to the case $R<\Lambda=\lambda^2/d$ ($\lambda$ is usual London penetration depth), and $d\ll \Lambda$, with $H$ near the lower critical field $H_{\rm {c1}}$. We further assume that $d$ is in the magnitude of $r_{c}$ (where $r_{c}$ is the radius of the vortex core), so that the sample can be approximated to 2D plane. We then choose cylindrical coordinates $(r,\theta,z)$ to simplify our calculations and we apply the method from Buzdin \cite{R10,R11} and Sobnack \cite{R9}. The restriction on $R$ implies that the screening effects of the superconducting currents ${\bf j_s}$ are suppressed: the whole disk can be thought of as soaking in the applied field $\bf H$, so that the local magnetic field ${\bf B}=\nabla \times {\bf A}$ is approximately the same inside and outside the sample, ${\bf B} \approx {\bf H}$. We formulate the problem at a temperature $T<T_{c}$ which is in superconducting state, where $T_{c}$ is the critical temperature.

\subsection{Ginzburg-Landau Model}
In this part, we will study the Gibbs Free energy \cite{R18, R24} in our small disk. We assume that the electrons in the superconducting state are in couple (cooper) pairs. From the $1^{st}$ quantization, the order-disorder of these electron pairs can be descibed as wave function \cite{R19, R20, R21}, $\Psi(x) = |\psi| e^{i\theta}$, where $|\psi|$ refers to the norm and $\theta$ is the phase of the wave function. We can think this as a vector with planar orientation. We also assume that the spatial variation of the $|\psi|$ to be very small. Under zero applied magnetic field, the free energy density \cite{R18, R24} of the system becomes 
\begin{equation}
g = g_{0} + \alpha |\Psi|^{2} + \frac{\beta}{2} |\Psi|^{4} + \texttt{higher order terms}, 
\end{equation}
where $\alpha$ and $\beta$ are parameters associated with superconducting material, and $g_{0}$ corresponds to the reference free energy density. This is the form of Talyor's approximation. If we further consider that the applied magnetic field $\bf H$ which is added on the system, two more significant terms need to concern. The first one is similar to the Hamiltonian of cooper pair under the electromagnetic field \cite{R18}, where the classical dynamic equation of the electrons pair is $$m\frac{d^2\bf {r}}{dt^2} = q{\bf E} + \frac{q}{c} {\bf v} \times {\bf B},$$ whereas $\bf E$ is the electric field which is zero in our case; $\bf v$ is the velocity of superconducting current and $q=2e$. Indeed, this is the kinetic enegry of electrons under the magnetic field. In the view of quantum mechanics, 
$$ \frac{1}{2m} \left({\bf p} - \frac{q}{c}{\bf A} \right)\Psi \cdot  \left({\bf p} - \frac{q}{c}{\bf A} \right)\Psi =  \frac{1}{2m}\left|  \left( \frac{\hbar}{i}\nabla - \frac{q}{c}{\bf A} \right) \Psi \right|^{2}. $$
$\bf A$ represents the magnetic vector potential and we can choose an arbitary gauage that is invariance. The second term is about the magnetic energy stored inside the system, $\frac{1}{8\pi} \left({\bf B} - {\bf H}  \right)^{2}$, where we define $\bf B$ as a local field generated by the superconducting system itslef due to the Meissner effect. This magnetic term refers to the normal energy in the sample. The total Gibbs free energy density under applied magnetic field $\bf H$, according to the Ginzburg-Landau theory, is hence,
\begin{eqnarray}
g &=& g_{0} + \alpha |\Psi|^{2} + \frac{\beta}{2} |\Psi|^{4} + \frac{1}{2m}\left|  \left( \frac{\hbar}{i}\nabla - \frac{q}{c}{\bf A} \right) \Psi \right|^{2} \nonumber\\
&+& \frac{1}{8\pi} \left({\bf B} - {\bf H}  \right)^{2}.
\end{eqnarray}
For a small disk, the screening is weak. The term of magnetic energy is small comparable to the kinetic energy and ${\bf H} \approx {\bf B}$ so that this magnetic term can be neglectable later.

Next, we need to minimize the given Gibbs free energy in order to find out the optimal solution. By variation principle, derivative with respect to the order parameter $\Psi$, we obtain the first equation of Ginzburg-Landau theory: 
\begin{equation}
\alpha \Psi + \beta \Psi |\Psi|^2 + \frac{1}{2m}\left( \frac{\hbar}{i}\nabla - \frac{q}{c}{\bf A}  \right)^2 \Psi = 0, 
\end{equation}
with boundary condition $\left( \frac{\hbar}{i}\nabla - \frac{q}{c}{\bf A} \right) \cdot {\bf n} = 0$ that normal component equals to zero, and only tangential component remains \cite{R15, R26}. The $1^{st}$ equation is analogy to the non-linear schrodinger equation which is time independent. While we take the differentiation with respect to $\bf A$ to the free energy, the $2^{nd}$ GL equation comes out. It corresponds to supperconducting density current $\bf j_{s}$:
\begin{equation}
{\bf j_{s}} = \frac{q \hbar}{2m i} \left(\Psi^*\nabla\Psi - \Psi\nabla\Psi^*\right)- \frac{q^2}{m c} \Psi^* \Psi {\bf A},
\end{equation}
which is similar to the quantum density current flow.

\subsection{Vortices inside the Small Thin Disk}
Up to this point, we are going to solve out the solution of vortices inside the thin disk. Suppose the order parameter is $\Psi(r) = |\Psi| e^{i\vartheta}$, as mentioned above. Then, we consider the gradient of order parameter $i\hbar \nabla \Psi$ which is associated with the quantum momentum, and related to the kinetic energy component. In the cylindrical coordinate, 
$$\nabla \Psi = \frac{\partial \Psi}{\partial r} \mbox{\boldmath$\hat r$} + \frac{1}{r} \frac{\partial \Psi}{\partial \theta} \mbox{\boldmath$\hat \theta$} +  \frac{\partial \Psi}{\partial z} \mbox{\boldmath$\hat z$}.$$
By the way, the disk can be treated as very thin so that we can approximate and then map it into 2-D cylindrical coordinate. We assume that the amplitude of $|\Psi|$ is amlost constant except at some singularities and the edge of the disk. It is on the other side to assume the spatial variation of $|\Psi|$ is very small and hence, 
\begin{equation}
\nabla \Psi \approx \frac{1}{r} \frac{\partial \Psi}{\partial \theta} \mbox{\boldmath$\hat \theta$} = \frac{i}{r} \nabla \vartheta \Psi \mbox{\boldmath$\hat \theta$}.
\end{equation}
Putting back into the Gibbs free energy equation, we obtain:
\begin{eqnarray}
g &\approx& g_{0} + \alpha |\Psi|^{2} + \frac{\beta}{2} |\Psi|^{4} + \frac{1}{2m}\left|  \left( \frac{\hbar}{r}\nabla \vartheta  - \frac{q}{c}{\bf A} \right) \Psi \right|^{2} \nonumber\\
\end{eqnarray}
What we need to do next is to handle the kinetic energy term. The gradient of the phase $ k = \nabla \vartheta$ represents the number of circulations of the vortex, and thus $k$ is the winding interger number.

Up to this point, we need to further discuss the physical mechanism of vortices in small type-II superconducting sample. If the applied field $H$ is smaller than the first critical field $H_{c1}$, the sample will tend to oppose $H$ (the Meissner effect). In the range of $H_{c1} \leq H \leq H_{c2}$, applied flux lines will penetrate into the disk and then vortices are being nucleated, with each vortex carrying a flux $\phi_v=l\phi_0$, 
where $l\in Z$ ($l>0$ for vortices and $l<0$ for anti-vortices) and $\phi_0=hc/q$ is the quantized fluxion. Typically, a core of radius $r_{c}$ in a vortex is in nano- magnitude of order $10^{-9}m$. On the other hand, outside the region of core radius, it more belongs superconducting state. Within the core region, it is a normal state rather than a superconducting state. Two more pararmeters are needed to clarify the range of normal and superconducting states: London penetration depth $\lambda$ and coherent length $\xi$. $\lambda$ measures how the distance relates to the applied magnetic field down to neglectable small inside the superconductor. And $\xi$ measures how the range of sample from normal state, affected by applied field, back to superconducting state.
We can imagine a vortex in the water pool. There is the circulation flow around the center which is usually a small hole. The hole can be regarded as the normal state while the flow refers to the super-current. In the thin mesoscopic disk, the circulating superconducting current around each vortex is approximated as ${\bf j}_v \propto \frac{1}{r} \hat \theta$.

Let's think about the kinetic energy which is contributed by the superconducting current, 
\begin{eqnarray}
KE &=&   \frac{1}{2m}\left|  \left( \frac{\hbar}{r}\nabla \vartheta  - \frac{q}{c}{\bf A} \right) \Psi \right|^{2} \nonumber\\
&=& \frac{\hbar ^2}{2m}\left|  \left( \frac{k}{r} - \frac{2 \pi}{\phi_0}{\bf A} \right) \Psi \right|^{2} \nonumber\\
&=& \frac{\hbar ^2}{2m} \left(\frac{2 \pi}{\phi_0}\right)^2  \left|  \left( \frac{\phi_0 k}{2 \pi r} - {\bf A} \right) \Psi \right|^{2}
\end{eqnarray}
It is obvious that ${\bf A_{v}} = {\phi_0 k}/{2 \pi r}$ is the magnetic potential generating by vortex. Now the Gibbs free energy expression is hence,
\begin{eqnarray}
g - g_{0} &=& \alpha |\Psi|^{2} + \frac{\beta}{2} |\Psi|^{4} \nonumber\\
           &+& \frac{\hbar ^2}{2m} \left(\frac{2 \pi}{\phi_0}\right)^2  \left|  \left( {\bf A_{v}} - {\bf A} \right) \Psi \right|^{2} 
\end{eqnarray}
After this, we have to handle the first two Talyor's terms of the free energy: 
$$
\alpha |\Psi|^{2} + \frac{\beta}{2} |\Psi|^{4} = \frac{\hbar ^2}{2m}\left( - \frac{1}{\xi^2} |\Psi|^2 + \frac{\beta}{2\xi^2 |\alpha|} |\Psi|^4 \right),
$$
where we define $\xi^2 = \hbar^2/2m|\alpha|$ \cite{R19, R24}. For the sake of minimization, $|\Psi_0|^2 = |\alpha|/\beta$, then the Taylor's terms will become: 
$$\frac{\hbar ^2 |\Psi_0|^2}{2m}\left( - \frac{1}{\xi^2} |\tilde{\Psi}|^2 + \frac{1}{2\xi^2} |\tilde{\Psi}|^4 \right),$$
where the dimensionless parameter is $\tilde{\Psi} = \Psi / |\Psi_0|$. By intergating both parts $\int {\bm d}^3r$ and we define free eenrgy from GL mdoel,
\begin{eqnarray}
G_{GL}  &=& \frac{\hbar ^2 |\tilde \Psi_0|^2}{2m} \int - \frac{1}{\xi^2} |\tilde \Psi|^2 + \frac{1}{2\xi^2} |\tilde \Psi|^4 
\nonumber\\
 &+& \left(\frac{2 \pi}{\phi_0}\right)^2  \left|  \left( {\bf A_{v}} - {\bf A} \right) \tilde{\Psi} \right|^{2} {\bm d}^3r, 
\end{eqnarray}
where $G = \int g {\bm d}^3r$. Besides, we can also approximate the applied magnetic potential ${\bf A_{\rm app}}$ to be very close to local magnetic potential $\bf A$ \cite{R10, R11}.

Similar the treatment to previous result \cite {R10, R11}, under certain temperature and magnetic field, the Gibbs free energy can be re-written in terms of number of vortices inside the small thin disk. We define $L= {-1,0,1}$ as a vortex at the center and $N$ as a number of vortices on a shell of the disk. If there is no vortices, say $L=0, N=0$, the normalized order parameter is $|\hat \Psi| \propto c$, where $c$ is a constant. Suppose the effect of the edge is neglectable. This parameter describes how much the superconducting and normal regions inside the system. While the square of order parameter represents the superconducting density of the system, $n_{s} \propto |\hat \Psi|^2 \propto c'$ (c' is constant). We only need to take into account where there are some vortices which will cause the superconducting region becomes normal state. Hence, we could assume that region (volume) of normal state created by a vortex can be approximated as $\pi \xi^2 d$, where $\xi$ refers to the coherent length of a vortex. The order parameter can be thus assumed as 
\begin{equation}
|\hat \Psi|^2 \propto c'(1-(\xi/R)^2(N+L)),
\end{equation}
where $N+L$ is the total number of vorticities.

\subsection{Vortex Shell Configurations on the Disk}
The first two terms of the free energy are related to the density of superconducting sample. So what we have to concern now is the kinetic energy term,  
\begin{eqnarray}
\frac{\hbar ^2 |\tilde \Psi_0|^2}{2m} \int \left(\frac{2 \pi}{\phi_0}\right)^2  \left|  \left( {\bf A_{v}} - {\bf A} \right) \Psi \right|^{2} {\bm d}^3r \approx \nonumber\\
\frac{\hbar ^2 |\tilde \Psi_0|^2}{2m} |\Psi|^ 2 \int \left(\frac{2 \pi}{\phi_0}\right)^2   \left( {\bf A_{v}} - {\bf A_{app}} \right)^{2} {\bm d}^3r
\end{eqnarray}
Here we assume the super-density $n_{s}\propto |\Psi|^2$ varies slowly in spatial and ${\bf A}\simeq {\bf A_{\rm app}}$. From this point, we can assume the local field be ${\bf B} \simeq {\bf H} = (0,0,H)$, whereas the magnetic potential vector can be chosen as ${\bf A} = (-yH/2,xH/2,0)$ - Landau gauage. More specifically,
\[ {\bf A} = \frac{1}{2}{\bf B} \times {\bf r}= \frac{1}{2} \left| \begin{array}{ccc}
{\hat i} & {\hat j} & {\hat k} \\
0 & 0 & H \\
x & y & z \end{array} \right|.\]

Now we need to choose the optimal postion of vortices. It is obtained by minimization of total free energy with respect to $r$. If there is only one vortex, the optimal position will be situated at the center of the disk. If there is more than one vortices, the vortices will at the position circulating around the center with radius $r = r_{i}$.   
The main concern is the interaction between vortices and local magnetic field. We can approximate this as analogy with electrostatics, by adding each vortex of flux $\phi_i$ at ${\bf r_i}$ in the disk, then there is an image anti-vortex of flux $-\phi_i$ at ${\bf r_i^\prime}=(R/r_i)^2\,{\bf r_i}$ beyond the disk and leads to 
\begin{equation}
{\bf A_v}=\sum_i \left [\Phi_i ({\bf r}-{\bf r_i})-\Phi_i ({\bf r}-{\bf r_i^\prime})\right ]\mbox{\boldmath$\hat{\theta}$},
\end{equation}
with $\Phi_i({\bf r})=\phi_i/(2\pi r)$. 
In our formulation, the vortices can be filled up to two shells. We assume that $N_1$ and $N_2$ be the number of vortices occupy on the two shells; where $L$ and $M$ are the number of quanta of vortices at the center and in the shell respectively. 

After some calculations, we can show that, at $T=0\,{\rm K}$, the Gibbs free energy of a configuration of $N_1$ vortices at $r_i=r_1$ and $N_2$ vortices at $r_j=r_2$, each of flux $\phi=M\phi_0$ in the shell, and a concentric vortex (anti-vortex) of flux $\phi=L\phi_0$ ($L>0$ for a vortex and $L<0$ for an anti-vortex) as
\begin{eqnarray}
G_{GL} \frac{2m}{\hbar ^2 |\Psi_0|^2 2 \pi d} = g_{GL}(L,N_1,N_2) =  \nonumber\\
 - \frac{R^2}{2\xi^2} |\tilde \Psi|^2 + \frac{R^2}{4\xi^2} |\tilde \Psi|^4 +  g_{Lond}(L,N_1,N_2)|\tilde \Psi|^{2}, 
\end{eqnarray}
whereas 
\begin{eqnarray}
g_{Lond}(L,N_1,N_2)&=& \frac{1}{4}h^2 + L^2\ln \frac{R}{r_{c}} - 2LN_1M\ln z_1 \nonumber\\
             &-& 2LN_2M\ln z_2 - Lh  + g^{\prime}(N_1,0)  \nonumber\\
             &+& g^{\prime}(N_2,0) + g_{12}(N_1,N_2), 
\end{eqnarray}
where $g = m  G /({\hbar ^2 |\Psi_0|^2 \pi d})$ is the dimensionless Gibbs free energy with reference, $h=H\pi R^2/\phi_0$ is the dimensionless applied field and $z_i = r_i/R ~(i=1,2)$ shows the optimal position. Furthermore, we discover that the $g_{Lond}$ is in fact the Gibbs free energy from London model.
$g_{Lond}^{\prime}(N_1,0)$ and $g_{Lond}^{\prime}(N_2,0)$ are the dimensionless free energies of $N_1$ and $N_2$ off-centre vortices respectively, with
\begin{eqnarray}
g^{\prime}(N_i,0)= N_iM^2\ln\frac{R}{r_{c}} - N_i(N_i-1)M^2\ln z_i \nonumber\\
     + N_iM^2\ln(1-z_i^2) - N_iMh(1-z_i^2) \nonumber\\
     + \frac{1}{2}N_iM^2\sum_{n=1}^{N_i-1}\ln\frac{1-2z_i^2\cos(2\pi n/N_i) + z_i^4}{4\sin^2(\pi n/N_i)}
\end{eqnarray}
for $(i=1,2)$ and, as is usual, we have introduced the core radius $r_{c}$ as a cutoff whenever ${\bf r_i}={\bf r_j}$. The interaction energy $g_{12}(N_1,N_2)$ of the $N_1$ vortices in the first shell (radius $r_1$) and the $N_2$ vortices in the second shell (radius $r_2$) is 
\begin{eqnarray}
g_{12}(N_1,N_2)=~~~~~~~~~~~~~~~~~ \nonumber\\
q^2\sum_{n,m}\ln \frac{1+z_1^2z_2^2-2z_1z_2\cos[\alpha+2\pi(n/N_1- m/N_2)]}{z_1^2+z_2^2-2z_1z_2\cos[\alpha+2\pi(n/N_1 - m/N_2)]},\nonumber\\
\end{eqnarray}
where $n\in [1,N_1-1], m\in [1,N_2-1]$ and $\alpha$ is the misalignment angle between vortices in the two shells.
Typically, we will take $L=M=1$ for the reason of mininization of energy.

\subsection{Entropy of Ordered Vortices associated with Temperature}
Next, we consider the arrangement of vortice inside the mesoscopic thin disk. The entropy \cite{R10} is introduced in our formulation. It measures how the order and disorder of the vortices placing on the disk. As the disk is in mesoscopic scale, the system is not allowed to put too many vortices on it. In our case, we assume there are two shells (rings) on the disk. The physical interpretation is that there are a number of arrangements of vortices on the shells, so that entropy exists when temperature is greater than zero but small than the critical limit. The general idea is thus
\begin{equation}
G(L,N_1,N_2,T) =  G(L,N_1,N_2) - TS, 
\end{equation}
where $S = k_{B} \ln (W)$ is total entropy of vortices and according to the Boltzmann statistics, $W$ is the number of configurations. At finite temperatures $T\neq 0\,{\rm K}$, one has to take into account the entropy $S$ associated with $N_1$ vortices at $r_i=r_1$ and $N_2$ vortices at $r_i=r_2$. This gives an additional term $-k_BT(\ln W_1 + \ln W_2)$, where $W_i=2\pi r_i/2N_i r_c ~(i=1,2)$, giving the dimensionless free energy as 
\begin{eqnarray}
g_{GL}(L,N_1,N_2,t)&=& g_{GL}(L,N_1,N_2)-t(2\ln \pi + 2\ln R/r_{c}\nonumber\\
&-&\ln N_1 - \ln N_2+\ln z_1+\ln z_2),
\end{eqnarray}
where $z_1=r_1/R$, $z_2=r_2/R$ and $t=mk_B T/(\hbar^2 |\Psi_0|^2 \pi d)$ is the dimensionless temperature. 
The magnetization $M$ of the disk follows from $\partial(G_{GL}+M\cdot H)/\partial H=0$ and this gives the reduced magnetization $m$ as $m(L,N_1,N_2)=-\partial g_{GL}(L,N_1,N_2,t)/\partial h$.

\section{Simulation Results and Analysis\protect}

\subsection{Discusion of cut-off Dimensionless Parameters}
We will go through the cut-off value of dimensionless parameters briefly. The normalized parameters $t$, $\xi$ and $\lambda$, and their effective ranges will be focused on. For the London equation, the penetration depth can be expressed as $\lambda = \sqrt {m c^2/4\pi n_s q^2}$, \cite{R18, R24}. The normalized temperature is $t=mk_B T/(\hbar^2 n_s \pi d)$, where $n_s = |\Psi_0|^2$. Let's consider the parament from GL formaulation, while $1/\pi n_s = 4\lambda^2 q^2 /m c^2$, $t=\lambda^2 k_{B} 16 \pi^2 T / \phi_0^2d$, whereas $\phi_0 = hc/q$. This exactly matches with what we find from the London theory. The effective $\lambda_{eff} \propto \lambda^2/d \approx 0.6 \sim 0.9R$ \cite{R19} is chosen as the weak screening effect by the small disk. It can be understood that magnetic field lines pass through and occupy in most regions of mesoscopic scale. The dimensionless ratio $\kappa = \lambda /\xi$ is an indicator to define type I and II superconductors. Here as we make use of effective penetration depth, we will assume the effective coherent length in parallel, such that $\xi_{eff} \propto \xi \lambda /d$. The physical interptation of $\xi_{eff}$ should be linking to the normal area state, $\pi \xi_{eff}^2$, created by a single vortex.

\subsection{Vortices inside Disk by Ginzburg-Landau Theory with Entropy}
For a given temperature $t$, we minimize $g_{GL}(L,N_1,N_2,t)$ with respect to $z_1$ and $z_2$ for a range of applied magnetic fields $h$, with different $L$, $N_1$ and $N_2$. Then the optimal Gibbs free energy is obtained and the transition for each state is considered.

\begin{figure}
\includegraphics[width=8cm,height=8cm]{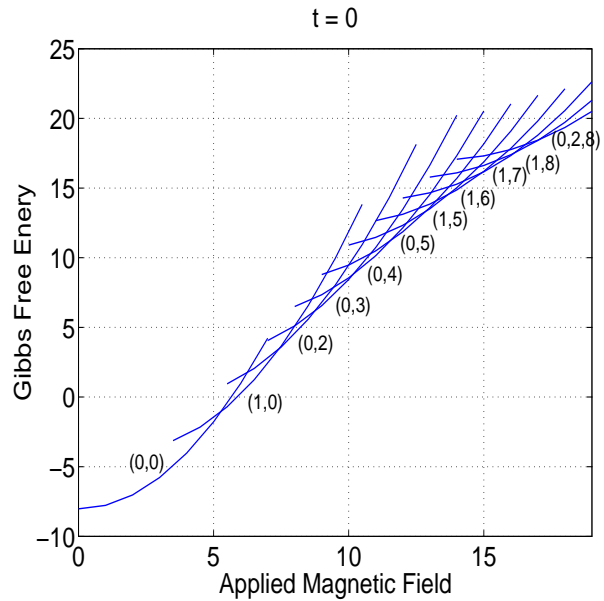}
\caption{\label{g, t=0} The free energy $g$ with respect to $h$ at $t=0$ is shown. The stable configurations are successsively $(0,0)\to(1,0)\to(0,2)\to(0,3)\to(0,4)\to(0,5)\to(1,5)\to(1,6)\to(1,7)\to(1,8)\to(0,2,8)$. The states $(L,N)=(1,5)$, $(1,6)$, $(1,7)$, $(1,8)$ and $(0,2,8)$ (corresponding to total flux $6\phi_0$, $7\phi_0$, $8\phi_0$, $9\phi_0$ and $10\phi_0$ respectively) are more stable than $(N,L)=(0,6)$, $(0,7)$, $(0,8)$, $(0,9)$ and $(0,10)$ or $(1,9)$.}
\end{figure}

\begin{figure}
\includegraphics[width=8cm,height=8cm]{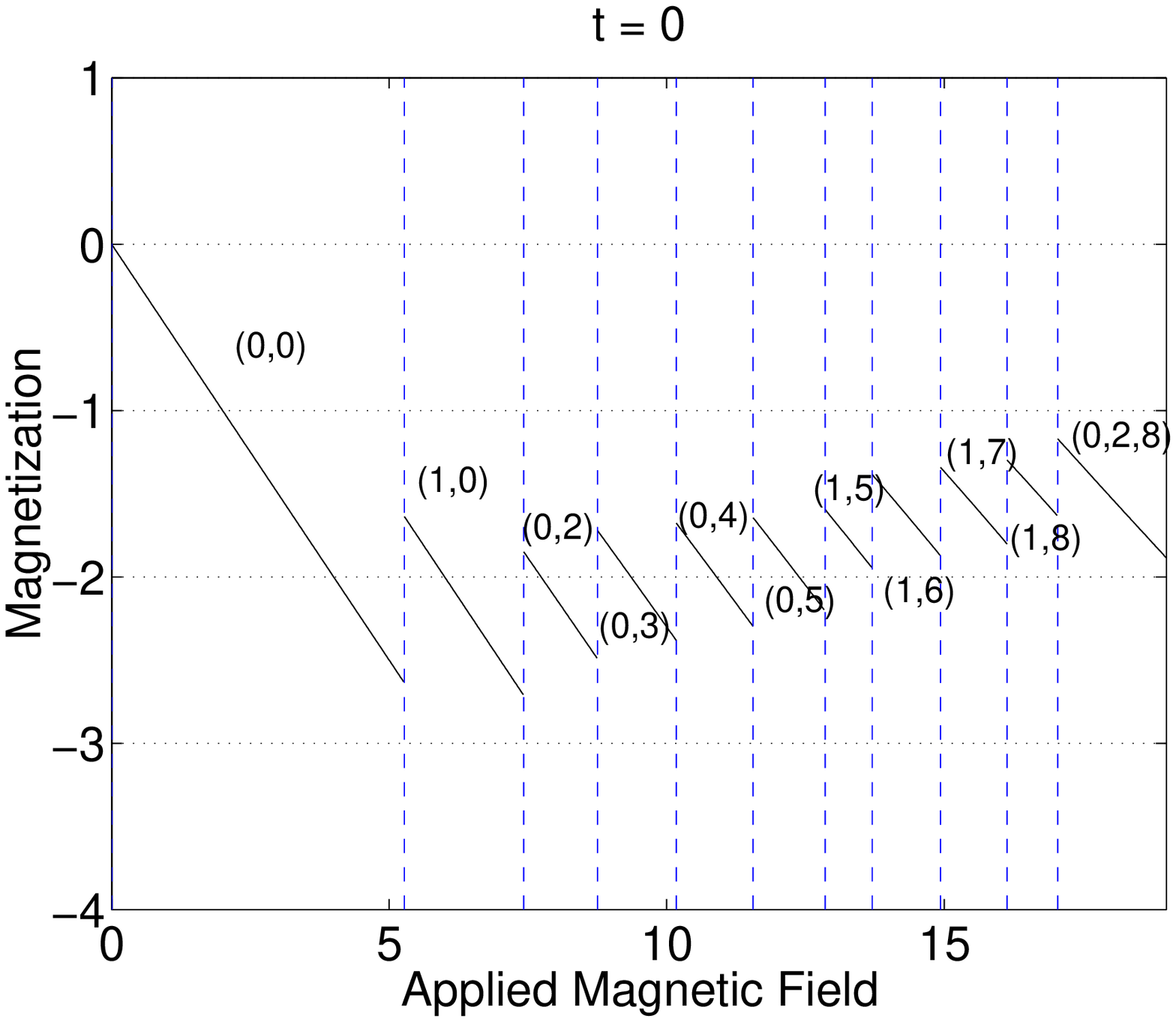}
\caption{\label{m, t=0} The magnetization $m$ with respect to $h$ at $t=0$. The stable configurations are $(0,0)\to(1,0)\to(0,2)\to(0,3)\to(0,4)\to(0,5)\to(1,5)\to(1,6)\to(1,7)\to(1,8)\to(0,2,8)$.}
\end{figure}

The critical temperature of Niobium is around $T_c=9.1\, {\rm K}$, so that in our dimensionless units, the critical temperature is $t_c=0.7$; while $t=0.14$ corresponds to the operative temperature $T=1.8\, {\rm K}$ at which the experiments were performed. Figures 1 and 2 show Gibbs free energy for $t=0$ and $t=0.14$ respectively. The parameters $d$, $R$, $\lambda$ are chosen so that our disk corresponds to the sample studied experimentally by Grigorieva {\it et al.} \cite{R16} and some approximations. [Nb, radius $R\approx 1.5 \sim 2.5\, \mu{\rm m}, \lambda \approx 90\, {\rm nm}, \xi \approx 15\, {\rm nm}, d\approx 0.1 \sim 0.3\xi$] 

As $h$ increases from zero (Figure 1), the free energy of the screening currents increases quadratically with $h$ as expected until the first critical field $h_1\sim 5.3$ is reached when a single vortex [$(L,N)=(1,0)$] penetrates the disk at the centre (see Fig.~1). This state persists until the second critical field $h_2\sim 7.4$ is reached at which the single center-vortex splits into two off-center vortices [$(0,2)$]. As $h$ is further increased, more off-center vortices nucleate on the first ring, forming successively a triangle, a square and a pentagon, until the sixth critical field  $h_6\sim 12.9$ is reached. Then it is energetically more favorable for the next vortex to nucleate at the center of the disk [$(1,5)$] than to form a hexagon of six off-center vortices with a vortex at each vertex. This result agrees with those of other studies \cite{R2,R7,R8,R11,R12} and is also analogous to the result of the study by Yarmchuck {\it el al.} \cite{R14} on the nucleation of vortices in superfluid $^4$He, which showed that a central vortex would appear in the system. As $h$ increases further, further off-center vortices enter the disk and nucleate on the first ring, the stable vortex states going through transitions $(1,5)\to(1,6)\to(1,7)\to(1,8)$, until the tenth critical field $\sim 17.0$ is reached when, instead of the tenth vortex nucleating to form state $(1,9)$, the vortices rearrange themselves to form the state $(L,N_1,N_2)=(0,2,8)$, with no vortex at the center of the disk, two vortices on the first ring and eight on the second. The entry of each additional vortex at successive critical fields is accompanied by a jump in the magnetization $m$ of the disc (shown in Figure 2).

\begin{figure}
\includegraphics[width=8cm,height=8cm]{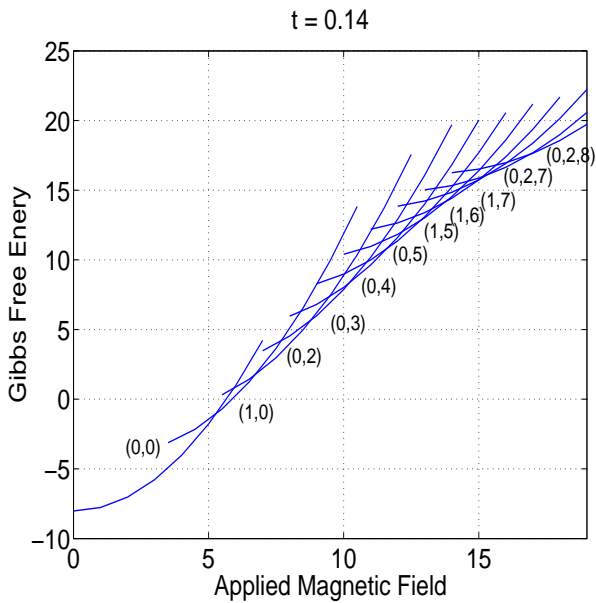}
\caption{\label{g, t=0.14} The free energy $g$ as a function of $h$ at $t=0.14$ ($T=1.8\,{\rm K}$). The stable vortex states and transitions are $(0,0)\to(1,0)\to(0,2)\to(0,3)\to(0,4)\to(0,5)\to(1,5)\to(1,6)\to(1,7)\to(0,2,7)\to(0,2,8)$ as $h$ increases (for the range of $h$ shown).}
\end{figure}

\begin{figure}
\includegraphics[width=8cm,height=8cm]{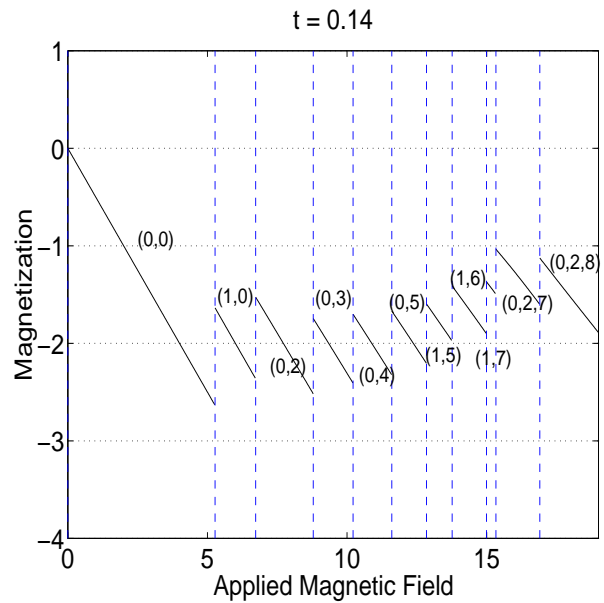}
\caption{\label{m, t=0.14} The magnetization $m$ as a function of $h$ at $t=0.14$ ($T=1.8\,{\rm K}$).The stable configurations are $(0,0)\to(1,0)\to(0,2)\to(0,3)\to(0,4)\to(0,5)\to(1,5)\to(1,6)\to(1,7)\to(0,2,7)\to(0,2,8)$.}
\end{figure}

At $t=0.14$, the Meissner state persists until the applied magnetic field reaches the first critical field $h_1\sim 5.3$ when a single vortex [$(L,N)=(1,0)$] nucleates at the center of the disk (see Figure 3). At the second critical field $h_2\sim 6.7$, the energetically favorable configuration (with total flux $2\phi_0$) is the state with two off-center single vortices $(L,N)=(0,2)$. As $h$ increases further, more vortices penetrate the disk, with the fluxiod state going successively through the transitions $(0,2)\to(0,3)\to(0,4)\to(0,5)\to(1,5)\to(1,6)\to(1,7)$ until the next critical field $h_8\approx 14.9$ is reached. Then an extra vortex enters the disk, but the stable vortex state with total flux $9\phi_0$ is the state $(0,2,7)$: no vortices at the centre of the disk, and the nine vortices form two rings, with two vortices on the inner ring and seven on the outer. As $h$ is further increased, a further vortex penetrates the disk and nucleates on the outer ring, forming the state $(0,2,8)$. These results are in direct agreement with the experimental observations of Grigorieva {\it et al.} \cite{R16}. Figure 4 shows the magnetization of the disk in which discontinuous jump is found.

\begin{figure}
\includegraphics[width=8cm,height=8cm]{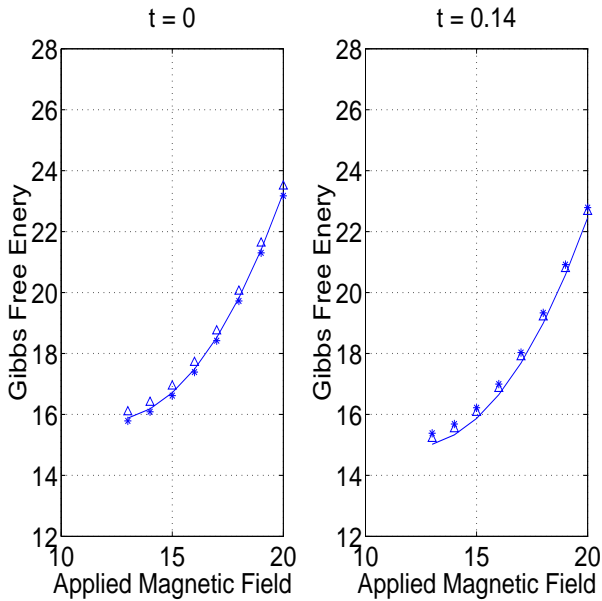}
\caption{\label{flux=9, t=0;>0} Free energy with total magnetic flux $=9\phi_0$. Blue, red and green lines represent the states $(1,8)$, $(0,2,7)$ and $(0,3,6)$ correspondingly. Left diagram reveals curves at $t = 0$ while right diagram is at $t  = 0.14$.}
\end{figure}

\begin{figure}
\includegraphics[width=8cm,height=8cm]{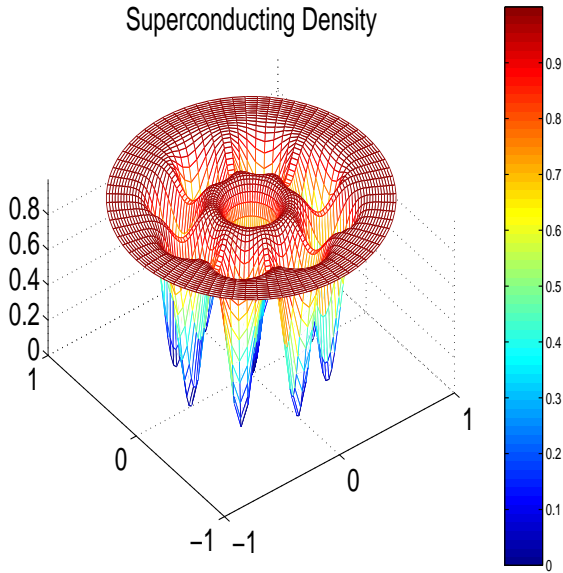}
\caption{\label{(1,8), t>0} Total flux $=9\phi_0$: State $(1,8)$ with one concentric vortex and eight off-center vortices.}
\end{figure}


\begin{figure}
\includegraphics[width=8cm,height=8cm]{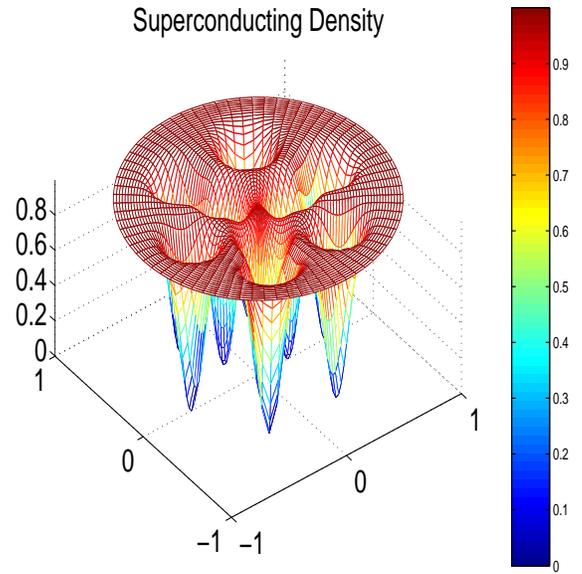}
\caption{\label{(0,3,6), t>0} Total flux $=9\phi_0$: State $(0,3,6)$ with three off center vortices in the $1^{st}$ ring and six off-center vortices in $2^{nd}$ ring.}
\end{figure}

Figure 5 extracts parts of figure 1 respectively, giving that Gibbs free energy with total flux $9\phi_0$. The solid line and triangle in Fig. 5 give the free energy of the vortex state $(0,2,7)$ and $(0,3,6)$ correspondingly (two shells configuration); while the star represents the state $(1,8)$, totally with flux $9\phi_0$. The most stable state at $t=0$ ($T=0$\,K) (Fig. 5 left) is the state $(1,8)$, with a central vortex and eight vortices on a shell. The next stable comfiguration is $(0,2,7)$ and the unstable one is $(0,3,6)$.

Whereas at $t=0.14$ ($T=1.8$\,K) (Fig. 5 right), the most stable state is the state $(0,2,7)$, with no vortex at the center of the disk, two vortices on the inner ring and seven on the outer ring. This result is in very good agreement with the experiments of Grigorieva {\it et al.} \cite{R16} (who found that at $T=1.8$\,K, the state $(1,8)$ was observed in only just a few cases, while the state $(0,2,7)$ was, by far, the most frequently observed state). From the studies of Baleus {\it et al.} \cite{R15}, they theoretically predicted only the state $(1,8)$ will be present. Figs 6-7 show the superconducting density with states $(1,8)$, $(0,2,7)$ and $(0,3,6)$ sequently. The color bar, representing the the norm of super- density, is shown from interval $[0,1]$. We can see that there are some holes inside the disk. Physically, those holes mean that the superconducting states are being destroyed by the field lines, which turns to normal state with zero magnitude. Magnitude with one regards as the superconducting region without destroy.

\begin{figure}
\includegraphics[width=8cm,height=8cm]{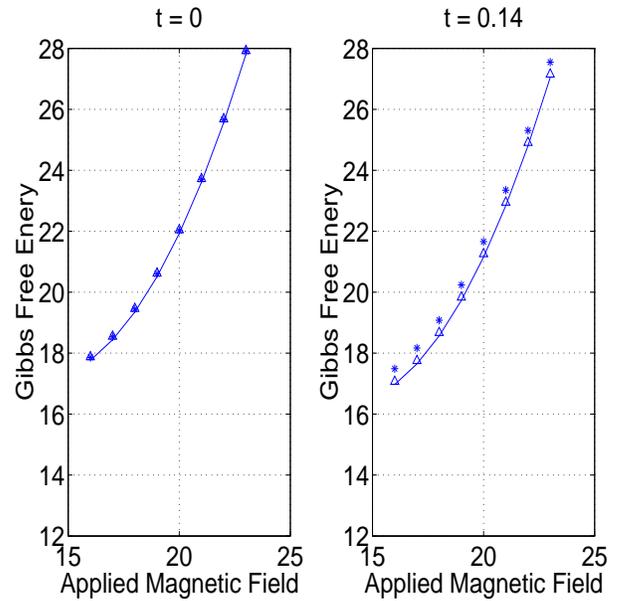}
\caption{\label{flux=10, t=0;>0} Free energy with total magnetic flux $=10\phi_0$. Blue, red and green lines represent the states $(1,9)$, $(0,2,8)$ and $(0,3,7)$ correspondingly. Left diagram reveals curves at $t = 0$ while right diagram is at $t  = 0.14$.}
\end{figure}


\begin{figure}
\includegraphics[width=8cm,height=8cm]{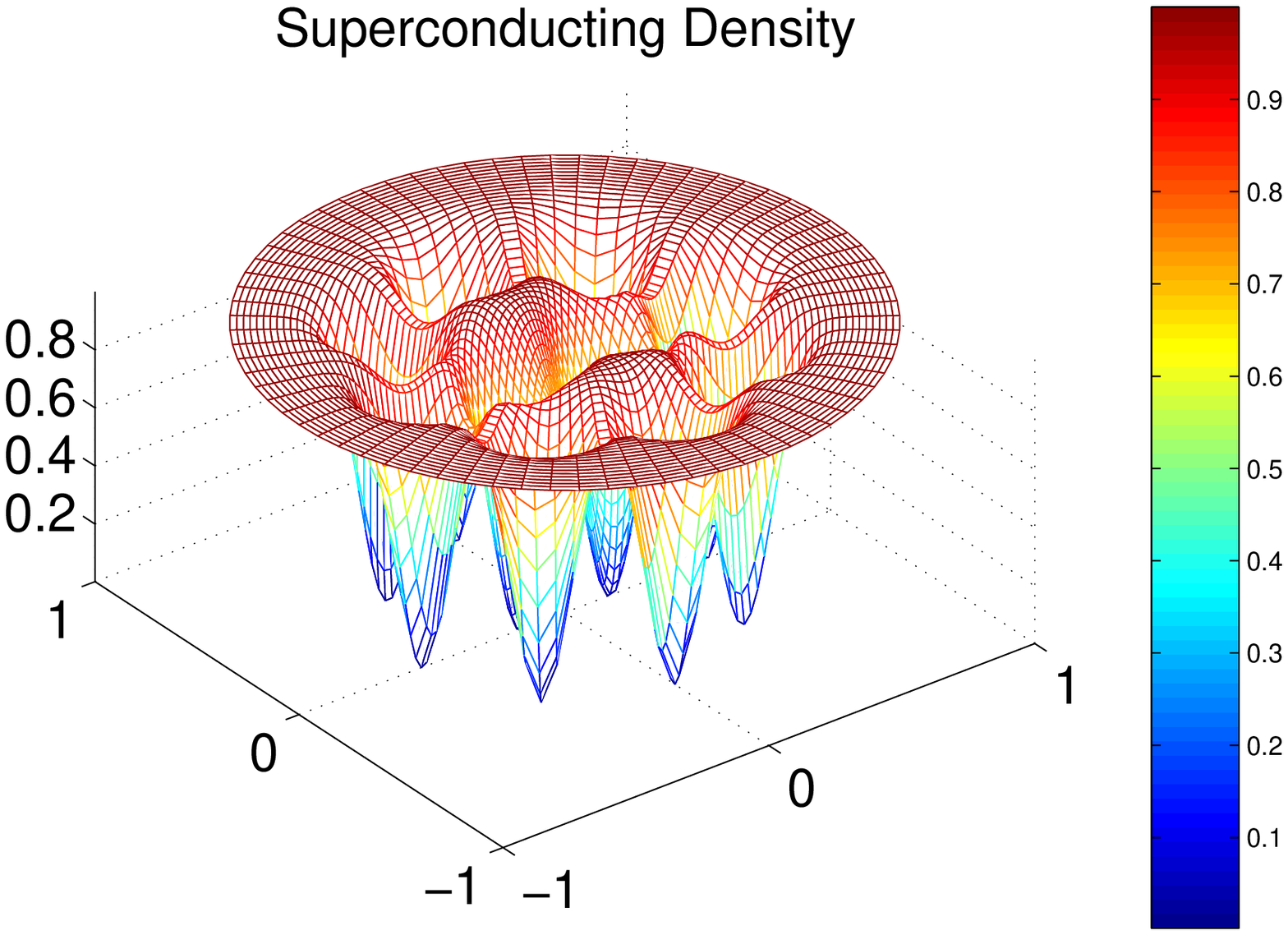}
\caption{\label{(0,2,8), t>0} Total flux $=10\phi_0$: State $(0,2,8)$ with two off center vortices in the $1^{st}$ ring and seven off-center vortices in $2^{nd}$ ring.}
\end{figure}

\begin{figure}
\includegraphics[width=8cm,height=8cm]{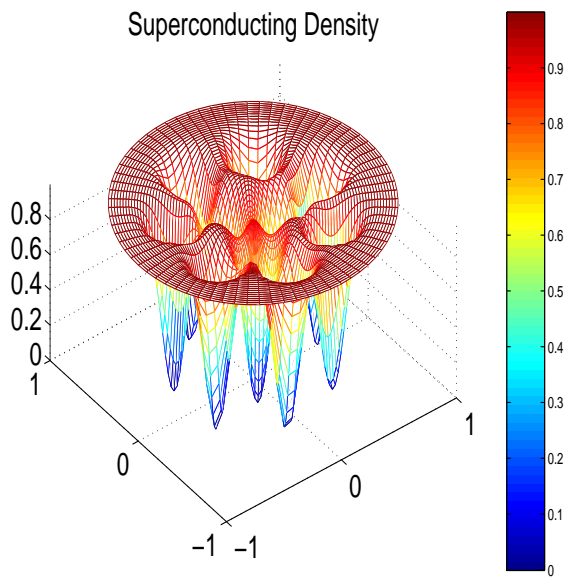}
\caption{\label{(0,3,7), t>0} Total flux $=10\phi_0$: State $(0,3,7)$ with two off center vortices in the $1^{st}$ ring and seven off-center vortices in $2^{nd}$ ring.}
\end{figure}

Figure 8 shows the corresponding result for vortex states with total flux $10\phi_0$. The triangle shows the free energy that depends on $h$ with state $(0,3,7)$, the star that represents the state of $(1,9)$ and the line represented by red colour is $(0,2,8)$. At $t=0$ (Fig. 8 left), the most stable vortex state is the state $(0,2,8)$, with states $(0,3,7)$ and $(1,9)$ having almost the same (slightly higher) energy. 

At $t=0.14$ ($T=1.8$\,K) (Fig. 8 right), the state $(0,2,8)$ is again the most stable state, with the other two states having higher energies. This result is in very good agreement with the experimental studies of Grigorieva {\it et al.} \cite{R16} who reported that the state $(0,2,8)$ was the most frequently observed state and that the state $(1,9)$ was never observed in their experiments. Baleus {\it et al.} \cite{R15}, on the other hand, predicted only the state $(1,9)$ for the state with total flux $10\phi_0$. Figs 9 and 10 are the density in states $(0,2,8)$ and $(0,3,7)$ respectively.

\subsection{Thermal Fluctuation of Quantized Vortices}
Experimentally\cite{R16}, it is found that the state $(0,2,7)$ with total flux $10\phi_0$ occurs more probable than the state $(1,8)$. The similar case appears in the subsequent state $(0,2,8)$ and $(0,3,7)$. Statistically, the occurance of vortex state $(0,2,7)$ is around 7 times more than $(1,8)$; while the appearance of $(0,2,8)$ is about 2-3 more than $(0,3,7)$. In order to explain the situation, we would like to reansonably approximate the probability of vortex state by the Maxwell-Boltzmann distribution. It is to say, $p \propto exp(-G_{GL}/k_BT)$, as the probability of certain configuration, where $G_i$ is the Gibbs free energy at the state $i$ and $k_B$ is the Boltzmann constant. For dimensionless, the probability expression can be written as 
\begin{equation}
p \propto exp(-g_{GL}/t), 
\end{equation}
where $g_{GL}$ is the normalized free energy.
Our approximation shows that the occurance of state $(0,2,7)$ is around 10 times more than $(1,8)$. We also predict that state $(0,3,6)$ will appear $2 \sim 3$ times more than $(1,8)$, in which it does not show in the experiment. Besides, the appearance of state $(0,2,8)$ is around $2 \sim 3$ times more than $(0,3,7)$. Both cases are consistent in the experiement.

\subsection{Gibbs Free Energy Compared with London Approximation}
\begin{figure}
\includegraphics[width=8cm,height=8cm]{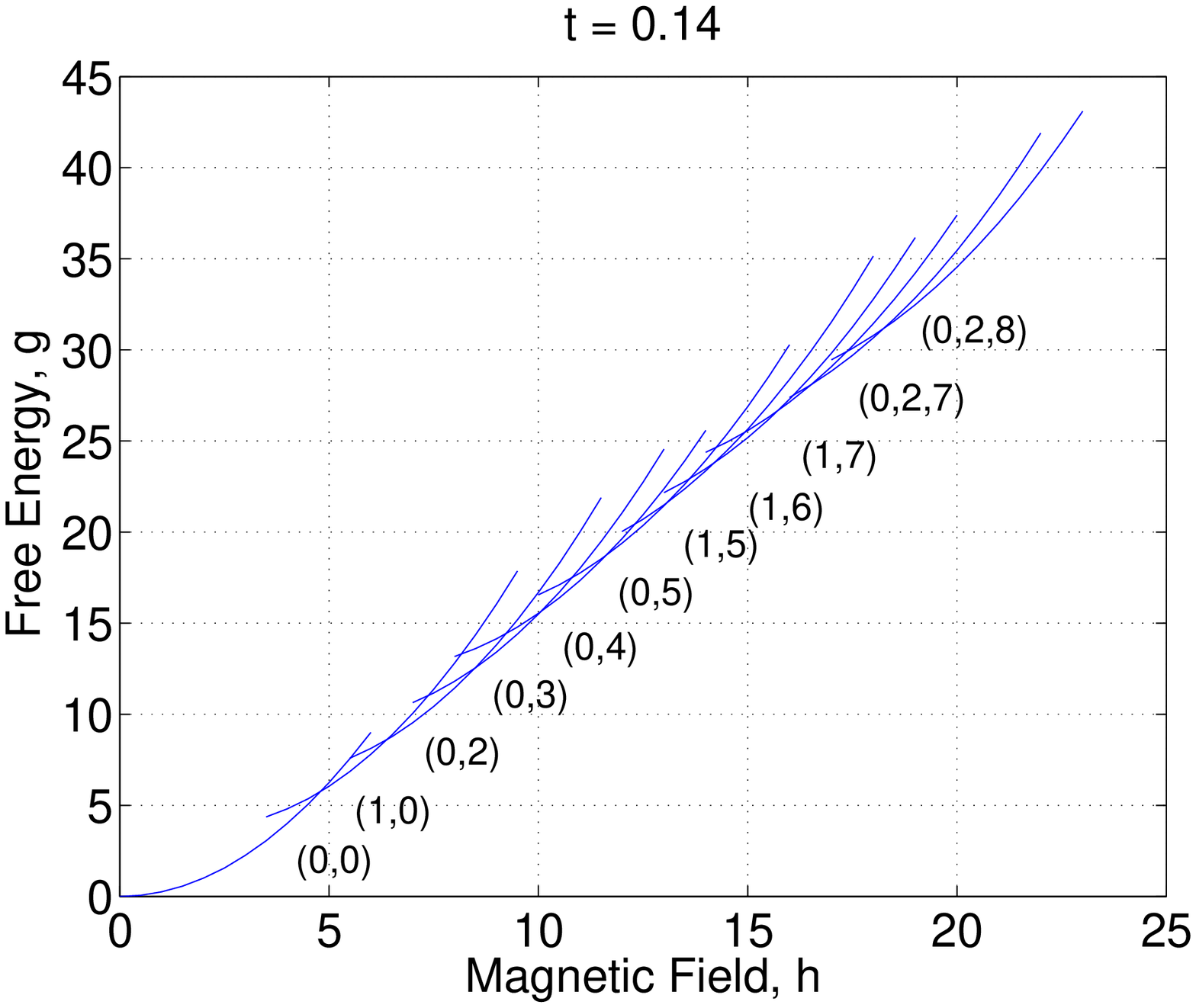}
\caption{\label{lond_t014} The free energy $g_{Lond}$ as a function of $h$ at $t=0.14$ ($T=1.8\,{\rm K}$).The stable configurations are $(0,0)\to(1,0)\to(0,2)\to(0,3)\to(0,4)\to(0,5)\to(1,5)\to(1,6)\to(1,7)\to(0,2,7)\to(0,2,8)$.}
\end{figure}

Finally, we compare the results between the Ginzburg-Landau and London theories. The Gibbs free energy derived from GL model can be approximated to the one by London model. From the general results, we found that both Gibbs free energies produce the same vortex state. One can find that the free energy from London model (fig.\ref{lond_t014}) has a higher level of magnitude than the one by GL model. It is due to the fact that the GL model also takes the supperconducting density $|\Psi|^2  = n_{s}$ into account. As the temperature or magnetic field increases further, it will undoubtedly destory part of the super-density which turns to normal state. Th free energy has therefore got a fall.

\section{Concluding Remarks}
We have formulated the Gibbs free energy by Ginzburg-Landau equations in a mesoscopic disk. A critical review of the early study of Sobnack and Kusmartsev \cite{R10}, which the London equations are applied, is made. We also introduce the idea of entropy into the Gibbs free energy. Inclusion of the temperature term $-TS$ (by taking into account the entropy $S$ associated with the non-center vortices) lowers the free energy of some of the vortex states and stabilizes them. Our results are in agreement with those of the recent experiments of Grigorieva and co-workers \cite{R16}. It is found that our results matches with those of Baleus {\it et al.} \cite{R15} in many cases, and there are only some states, $(1,8)$ and $(1,9)$ which show some differences. Those different states are also predicted by our modified formulation which coincide in the experiments \cite{R16}. A possible reason for the disagreement between the theory of Baleus {\it et al.} \cite{R15} and ours is that the experiments are performed at finite temperatures $T\neq0$\,K, whereas the study of Baleus {\it et al.} \cite{R15} concern the circumstance at $T=0$\,K. The free energies by London and Ginzburg-Landau equations have been investigated. We find that free energy formulated by Ginzburg-Landau theory provides a lower level of energy. Physically, it means that the ingredient of superconducting density is taken into account. More flux lines will break the order symmetry and some regions will transit to normal. Broadly speaking, both free energies can be obtained the same configuration of vortices in small disck. It is shown that the free energy by Ginzburg-Landau theory can be reasonably approximated to London theory.

\bibliography{apssamp}

\end{document}